\documentclass{osa-article}

\journal{osac}

\articletype{Research Article}

\begin{document}

\title{Nanophotonic Media for Artificial Neural Inference}

\author{Erfan Khoram,\authormark{1} Ang Chen,\authormark{1}, Dianjing Liu\authormark{1}, Lei Ying\authormark{1}, Qiqi Wang\authormark{2}, Ming Yuan\authormark{3} Zongfu Yu\authormark{1,*}}

\address{\authormark{1}Department Electrical And Computer Engineering, University of Wisconsin Madison-Madison,WI53706,USA\\
\authormark{2}Department of Aeronautics and Astronautics, Massachusetts Institute of Technology,Cambridge,MA02139,USA\\
\authormark{3}Department of Statistics, Columbia University,New York,NY10027,USA}

\email{\authormark{*}zyu54@wisc.edu} 



\begin{abstract}
We show optical waves passing through a nanophotonic medium can perform artificial neural computing. Complex information, is encoded in the wave front of an input light. The medium transforms the wave front to realize sophisticated computing tasks such as image recognition. At the output, the optical energy is concentrated to well-defined locations, which for example can be interpreted as the identity of the object in the image. These computing media can be as small as tens of wavelengths and offer ultra-high computing density. They exploit sub-wavelength scatterers to realize complex input output mapping beyond the capabilities of traditional nanophotonic devices. 
\end{abstract}

\section{Introduction}
Artificial neural networks (ANN) have shown exciting potential in a wide range of applications, but they also require ever-increasing computing power. This has prompted an effort to search for alternative computing methods that are faster and more energy efficient. One interesting approach is optical neural computing \cite{shen2017deep,hermans2015trainable,hughes2018training,skinner1995neural,prucnal2017neuromorphic,chen2016asp,bueno2018reinforcement}. This analog computing method can be passive, with minimal energy consumption, and more importantly, its intrinsic parallelism can greatly accelerate computing speed. 

Most optical neural computing follow the architecture of digital ANNs, using a layered feed-forward network as shown in Fig. 1a. Free-space diffraction\cite{lin2018all,skinner1995neural} or integrated waveguides\cite{shen2017deep,hughes2018training,hermans2015towards} are used as the connections between layered activation units. Similar digital signals in ANN, optical signals pass through optical networks in the forward direction once (light reflection propagating in the backward direction is avoided or neglected). However, it is the reflection that provides the feedback mechanism which gives rise to rich wave physics. It holds the key to the miniaturization of optical devices such as laser cavities \cite{park2004electrically}, photonic crystals \cite{joannopoulos2011photonic}, meta-materials \cite{cai2009optical}, and ultra-compact beam splitters \cite{shen2015integrated,piggott2017fabrication,su2017inverse}. Here we show that by leveraging optical reflection, it is possible to go beyond the paradigm of layered feed-forward networks to realize artificial neural computing in a continuous and layer-free fashion. Fig.1b shows the proposed Nanophotonic Neural Medium(NNM). An optical signal enters from the left and the output is the energy distribution on the right side of the medium. Computation is performed by a host material, such as $SiO_2$, with numerous inclusions. The inclusions can be air holes, or any other material with an index different from that of the host medium. These inclusions strongly scatter light in both the forward and backward directions. The scattering spatially mixes the input light, rendering it a counterpart to linear matrix multiplication (Fig.1c) in a digital ANN. The locations and shapes of inclusions are the equivalent of weight parameters in digital ANNs, and their sizes are typically sub-wavelength. The nonlinear operation can be realized via inclusions made of dye semiconductor or graphene saturable absorbers, where they perform distributed nonlinear activation. These nonlinearities are designed with the Rectified Linear Units (ReLU) in mind\cite{nair2010rectified}(Fig.1d).  

\begin{figure}[ht]
\centering
\includegraphics[scale = 1]{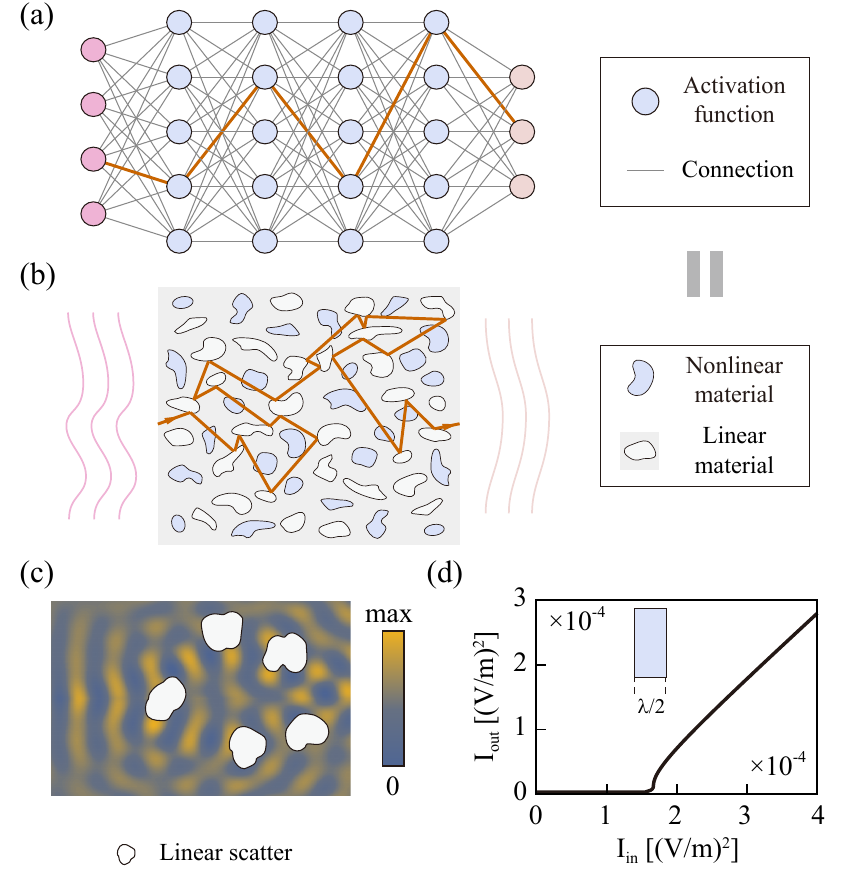}
\caption{(a) A conventional ANN architecture where the information propagates only in the forward direction(green arrow). (b) Proposed Nanophotonic Neural Medium (NNM). Passive neural computing is performed by light passing through nanostructured medium with both linear and nonlinear scatterers. (c) Full-wave simulation of light scattered by nanostructures, which spatially redistribute the optical energy to different directions. (d) The output intensity of light with wavelength $\lambda$, passing through the designed nonlinear material with a thickness of $\lambda/2$. It is a nonlinear function of the incident wave intensity. This material is used as nonlinear activation as indicated by light blue color. 
}
\label{fig:compar}
\end{figure}

Fig.2 shows a NNM in action, where a two-dimensional (2D) medium is trained to recognize gray-scale handwritten digits. The dataset contains 5,000 different images, representative ones of which are shown in Fig.2a. Each time, one image, represented by $20\times20$ pixels, is converted to a vector, and then encoded as the spatial intensity of input light incident on the left. Inside the NNM, nanostructures create strong interferences and light is guided toward one of ten output locations depending on the digit that the image represents, where the output with the highest share of energy intensity is categorized as the inferred class. Fig.2b shows the fields created by two different hand-written $2$ digits. Because of different shapes, the field patterns created by these two images are quite different but both lead to the same hot spot at the output, which correctly identifies the identity information as the number $2$. As another example, Fig.2c shows the case of two handwritten $8$ digits that result in another hot spot. Here, the field is simulated by solving a nonlinear wave equation using Finite-Difference Frequency-Domain (FDFD\cite{maxwellfdfd-webpage}) method. The size of the NNM is $80\lambda$ by $20\lambda$, where $\lambda$ is the wavelength of light used to carry and process the information. The average recognition accuracy reaches over $79\%$ for a test set made up of 1,000 images. The limited reported accuracy is due to the heavy constraints we set during the optimization for fabrication concerns. These constraints keep the medium dense, where it would have been otherwise made up of sparse sections of air and $SiO_2$. By relaxing these requirement or using larger medium sizes, the accuracy can be further improved. 

\begin{figure}[ht]
\centering
\includegraphics[scale=1]{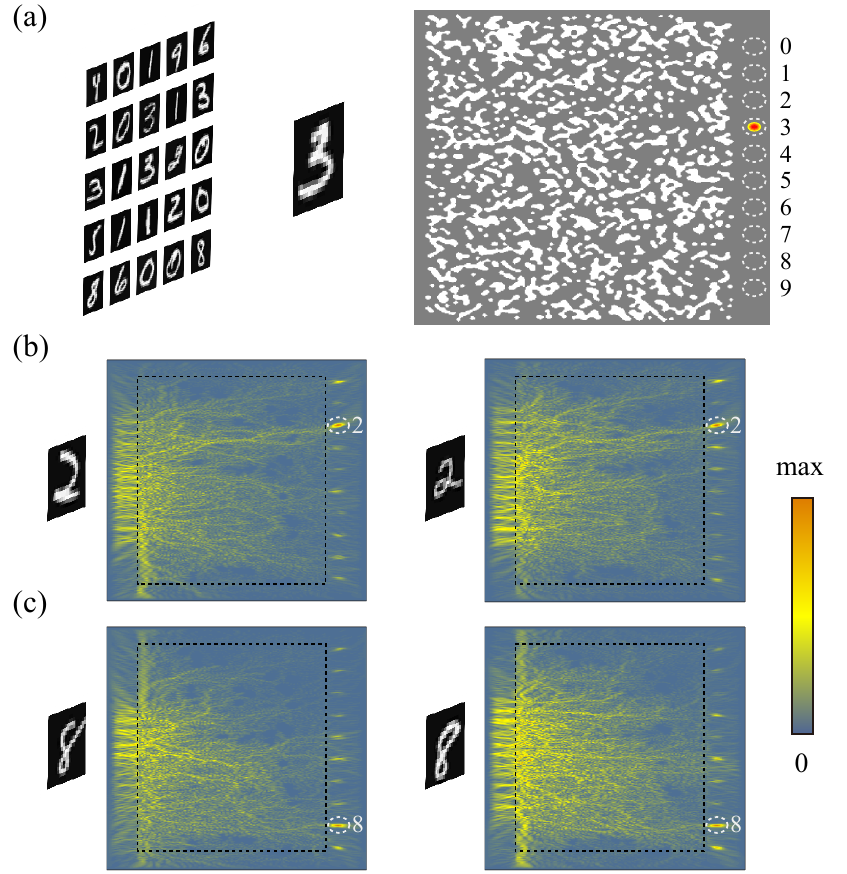}
\caption{(a) NNM trained to recognize handwritten digits. The input wave encodes the image as the intensity distribution. On the right side of NNM, the optical energy concentrate to different locations depending on the image's classification labels. (b)Two samples of the digit $2$ and their optical fields inside NNM. As it can be seen, although the field distributions differ for the images of the same digit, they are classified as the same digit. (c) the same as (b) but for two samples of the digit $8$. Also, in both (b) and (c), the boundaries of the trained medium have been shown with black borderlines.(see Visualization 1)} 
\label{fig:fields}
\end{figure}

NNM can provide ultra-high computing density by tapping into sub-wavelength features. In theory, the number of weight parameters is infinite: every atom in this medium can be varied to influence the wave propagation. In practice, a change below 10 nm would be considered too challenging for fabrication. Even at this scale, the potential number of weights exceeds 10 billion parameters per square millimeter for a 2D implementation. This is much greater computing density than both free-space \cite{caulfield1989optical,lin2018all} and on-chip optical neural networks \cite{shen2017deep,hughes2018training}. In addition, NNM has a few other attractive features. It has stronger expressive power than layered optical networks. In fact, layered networks are a subset of NNM, as a medium can be shaped into connected waveguides as a layered network. Furthermore, it does not have the issue of diminishing gradients in deep neural networks. Maxwell's equations, as the governing principle, guarantee that the underlying linear operation is always unitary, which does not have diminishing or exploding gradients\cite{pmlr-v70-jing17a}. Lastly, NNM does not have to follow any specific geometry, and thus it can be easily shaped and integrated into existing vision or communication devices as the first step of optical preprocessing.

We now discuss the training of NNM. Although, one could envision in-situ training of NNM using tunable optical materials \cite{hughes2018training}, here we focus on training in the digital domain and use NNM only for inference. The underlying dynamics of the NNM are governed by the nonlinear Maxwell's equations, which, in the frequency domain, can be written as
\begin{equation}\label{eq_1}
\begin{gathered}
L_{(r,E_{(r)})}E_{(r)}=-i\omega J_{(r)}
\end{gathered}
\end{equation}
where $L_{(r,E_{(r)})}=(\nabla\times\nabla\times)/\mu-\omega^2\varepsilon_{(r,E_{(r)})}$, and $\mu$ and $\varepsilon$ are the permeability and permittivity. $J$ is the current source density which represents the spatial profile of the input light and is only non-zero on the left side of the medium. Waveguide modes or plane waves can also be used as the input, which are also implemented as current sources in numerical simulation. For a classification problem, the probability of the $i^{th}$ class label is given by $h_{i}=(\int_{dr} |E(r)|^2R_i(r)) / (\sum_{i=1}^{10}\int_{dr} |E(r)|^2R_i(r))$, which represents the percentage of energy at the $i^{th}$ receiver relative to the total optical energy that reaches all receivers. Here the profile function $R_i(r)$ defines the location of receivers, and is only non-zero at the position of the $i^{th}$ receiver. The training is performed by optimizing the dielectric constant $\varepsilon(r,E)$ similar to how weight parameters are trained in traditional neural networks. 
The cost function $C$ is defined by the cross entropy between the output vector $\textbf{h}$ and the ground truth $\textbf{y}$. 
\begin{equation} \label{eq_2}
\begin{gathered}
C=-\sum_{i=1}^{10}y_{i} \log(h_{i})+(1-y_{i})\log(1-h_{i})\\
\end{gathered}
\end{equation}
The ground truth $\textbf{y}$ is a one-hot vector. Digit $8$ is represented as $\textbf{y}=(0,0,0,0,0,0,0,0,1,0)$, for instance. The gradient of the cost function with respect to the dielectric constant $\varepsilon$ can be calculated point by point. For example, one could assess the effect of changing $\varepsilon$ at one spatial point; the change is only kept if the loss function decreases. This method has achieved remarkable success in simple photonic devices \cite{shen2015integrated}. However, each gradient calculation requires solving full-wave nonlinear Maxwell's equations. It is prohibitively costly for NNM, which could easily have millions of gradients. Here, we use Adjoint State Method(ASM) to compute all gradients in one step:
\begin{equation}\label{eq_3}
\begin{gathered}
\frac{dC}{d\varepsilon_{(r)}}=-2\omega^2\,Real\{\lambda_{(r)} E_{(r)}\}
\end{gathered}
\end{equation}
Here $\lambda{(r)}$ is a Lagrangian multiplier, which is the solution to the adjoint equation (Eq.4), in which the electric field $E(r)$ is obtained by solving Eq.1. The adjoint equation here is slightly more involved than what is generally used in inverse design, and this is due to the fact that nonlinear behavior is included in our dynamics. A similar derivation for nonlinear adjoint equation is done in \cite{hughes2018adjoint}.

\begin{equation}\label{eq_4}
\begin{gathered}
\frac{\partial C}{\partial E_{(r)}}+\lambda_{(r)}(L_{(r,E_{(r)})}+\frac{\partial L_{(r,E_{(r)})}}{\partial E_{(r)}}E_{(r)})+\overline{\lambda_{(r)}}(\frac{\partial \overline{L_{(r,E_{(r)})}}}{\partial E_{(r)}}\overline{E_{(r)}})=0
\end{gathered}
\end{equation} 

The training process, as illustrated in Fig.3a, minimizes the summation of the cost functions $C$ for all training instances through Stochastic Gradient Descent(SGD). The process starts with one input image as the light source, for which we solve the nonlinear Maxwell's equations in an iterative process (pink block in Fig.3a). The initial field is set to be random $E_0(r)$, which allows us to calculate the dielectric constant $\varepsilon(r,E_0(r))$. Then FDFD simulation is used to solve Eq.1, and the resulting electric field $E_1(r)$ is then used to update the dielectric constant. This iteration continues until the field converges. The next step is to compute the gradient based on Eq.3. Once the structural change is updated, the training of this instance is finished. 
\begin{figure}[ht]
\centering
\includegraphics[scale=1]{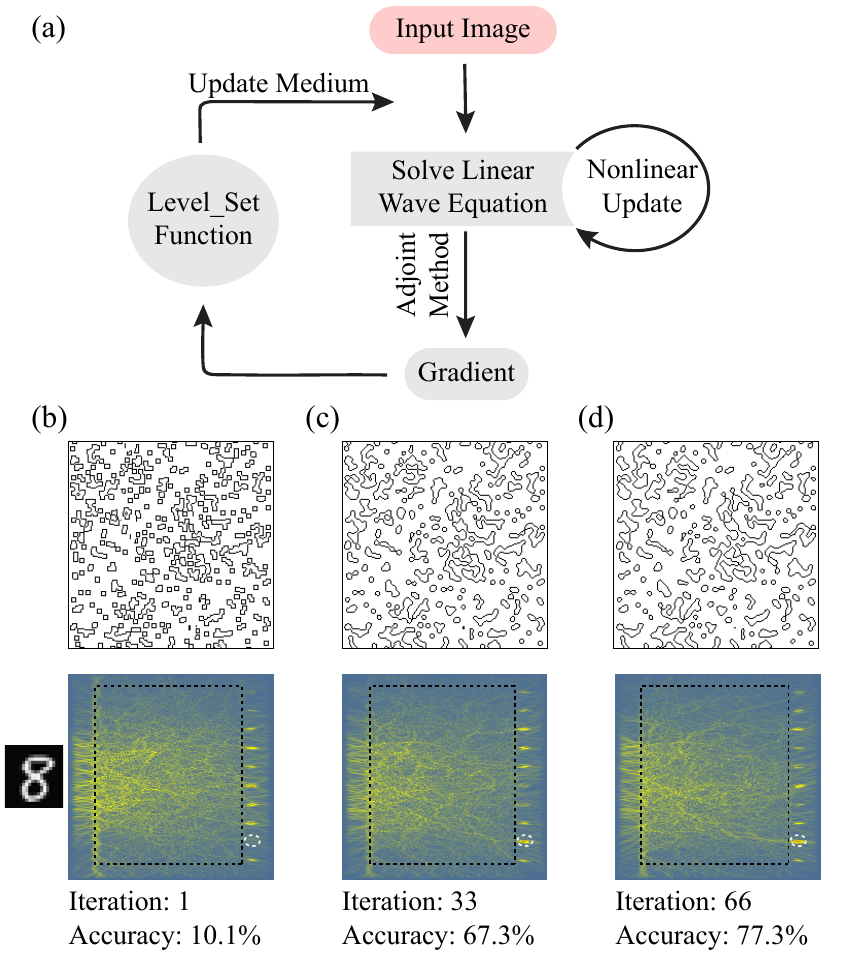}
\caption{(a)The training starts by encoding an image as a vector of current source densities in the FDFD simulation.  This step is followed by an iterative process to solve for the electric field in a nonlinear medium. Next we use the adjoint state method to calculate the gradient, which is then used to update the level-set function and consequently, the medium itself. In batch training, we sum the cost functions calculated for different images in the same batch and compute the gradients. (b)-(d) show an NNM in training after 1, 33, and 66 training iterations respectively. At each step, the boundary between the host material and the inclusions is shown, along with the field distribution for the same randomly selected digit $8$. Also the accuracy of the medium on the test set can be seen for that particular stage in training.}
\label{fig:evolution}
\end{figure}

The above process is repeated again, but for the next image in the training queue, instead of the same image. This gradient descent process is stochastic, which is quite different from typical use of ASM in nanophotonics \cite{piggott2017fabrication,su2017inverse} where gradient descent is performed repeatedly for very few inputs until the loss function converges. In these traditional optimizations, the device needs to only function for those few specific inputs. If such processes were used here, the medium would do extremely well for particular images but fail to generalize and recognize other images. 

The gradient descent process treats the dielectric constant as a continuous variable, but in practice, its value is discrete, depending on the material used at the location. For example, in the case of a medium with $SiO_2$ host material and linear $air$ inclusions, the dielectric constants can either be 2.16 or 1. Discrete variables remain effective for neural computing \cite{courbariaux2016binarized}. Here, we need to take special care to further constrain the optimization process. This is done by using a level set function \cite{li2010distance}, where each of the two materials (host material and the linear inclusion material), is assigned to each of the two levels in the level-set function $\phi(r)$ similar to ref \cite{piggott2017fabrication,su2017inverse}.

\begin{equation} \label{eq5}
\varepsilon(r)=\left\{
                \begin{array}{ll}
                  \varepsilon_{SiO_{2}}\quad\phi(r)<0\\
                  \varepsilon_{Air}\quad\phi(r)>0
                \end{array}
              \right.
\end{equation}
The training starts with randomly distributed inclusions, both linear and nonlinear, throughout the host medium. The boundaries between two materials evolve in the training. Specifically, the level set function is updated by -$v(r)|\nabla\phi|$, where $v(r)$ is the gradient calculated by ASM and $|\nabla\phi|$ indicates the boundary between the two constituent materials. Therefore, at each step, this method essentially decides whether any point on the boundary should be switched from one material to the other. Nonlinear sections perform the activation function and their location and shape are fixed in this optimization. They could also be optimized, which would be equivalent to optimizing structural hyperparameters in layered neural networks \cite{bergstra2012random,snoek2012practical,saxena2016convolutional}.

As a specific example, we now discuss the training of the 2D medium shown in Fig.2. The structural evolution is shown in Fig.3b-d during the training. We start by randomly seeding the domain with dense but small inclusions. As the training progresses, the inclusions move and merge, eventually converging. The recognition accuracy for both training and test group improve during this process.

Next, we show another example based on a three-dimensional(3D) medium, whose size is $4\lambda\times 4\lambda\times 6\lambda$. The inputs can be an image projected on the top surface of the medium. For example, we use a plane wave to illuminate a mask with its opening shaped into a hand-written digit as shown in Fig.4(Movie S2 shows how the energy distribution on the output evolves as a handwritten digit gradually emerges as the input). Fabricating 3D inclusions is generally difficult, but it is much easier to tune the permittivity of materials using direct laser writing \cite{marshall2009laser}. Thus, here we allow the dielectric constant to vary continuously. To save on computational resources, we allow $5\%$ variation. In experimental realization, a smaller variation range can always be compensated for by using larger media. The 3D trained NNM had an accuracy of about $84\%$ for the test set; the confusion matrix is shown in Fig.4b. The better performance in comparison with the 2D implementation is due to a higher degree of freedom we allow the dielectric constant to have.

\begin{figure}[ht]
\centering
\includegraphics[scale = 1]{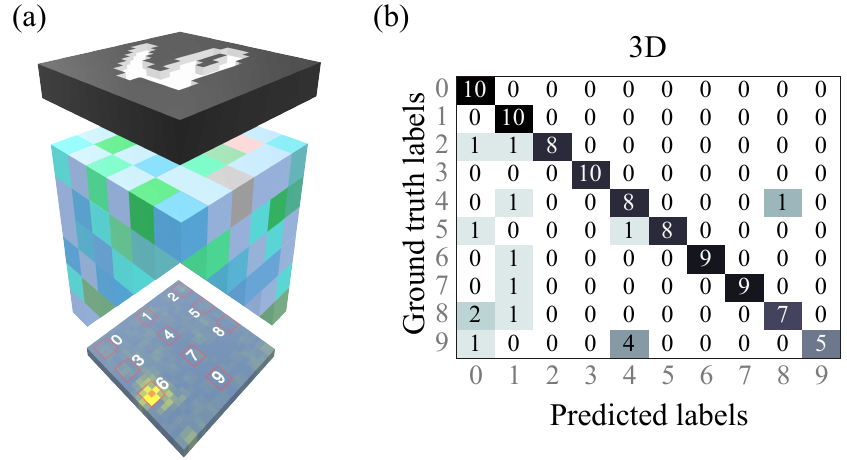}
\caption{(a) 3D nanophotonic neural medium case. Different colors illustrate varying values of permittivity. The input image is projected onto the top surface. Computing is performed while the wave propagates through the 3D medium. The field distribution on the bottom surface is used to recognize the image. Full-wave simulation shows the optical energy is concentrated on the location with correct class label, in this case 6.(b) The confusion matrix. The rows on the matrix show true labels of the images that have been presented as input, and the columns depict the labels that the medium has classified each input. Therefore the diagonal elements show the number of correct classifications out of every ten samples.(see Visualization 2)}
\label{fig:shemat3d}
\end{figure}

\section{Supplemental Materials}
\renewcommand{\theequation}{S\arabic{equation}}
\setcounter{equation}{0}
\newcommand*\mycommand[1]{\texttt{\emph{#1}}}
\renewcommand{\thefigure}{S\arabic{figure}}

\subsection{Gradient Derivation}

In this section we will go through the derivation of the gradient of the cost function for the case of a classification problem with mini-batch gradient descent. We start with the the electromagnetic wave equation in the frequency domain:

\renewcommand{\theequation}{S\arabic{equation}}
\setcounter{equation}{0}

\begin{equation}\label{eq_s1}
\begin{gathered}
L_{(r,\omega)}E_{(r)}=-i\omega J_{(r)}\\
L_{(r,\omega)}=(\nabla\times(\frac{1}{\mu})\times\nabla\times)-\omega^2\varepsilon_{(r)}
\end{gathered}
\end{equation}

Here $J$ is the current source density which corresponds to the values of the image pixels in our work. This equation is written for linear materials; For nonlinear materials, $\varepsilon$ becomes a function of the electric field at that point. We will elaborate upon this point as we derive the gradient.

The receivers on the output side of the structure measure the electromagnetic energy in their position. To model the light absorption in receivers, we use a Gaussian function to define the profile of the location of the $i^{th}$ output.

\begin{equation}\label{eq_s2}
R_{i(r)}=e^{-\frac{||r-r_{i}^{rec}||}{2\sigma^2}}
\end{equation}

where $r_{i}^{rec}$ is the location of the $i^{th}$ receiver while $\sigma$ represents the spatial span of the receivers. The energy inside the receiver is  

\begin{equation}\label{eq_s3}
o_i=\frac{\varepsilon_{air}}{2}\int ||E_{(r)}||^2 R_{i(r)}\,dr 
\end{equation}

In this equation, $\varepsilon_{air}$ is set as the permittivity of air, as the receivers are located outside the medium. Then the cross entropy cost function for a batch of size $m$  can be written as 

\begin{equation} \label{eq_s4}
\begin{gathered}
C=-\frac{1}{m}\sum_{j=1}^{m}\sum_{i=1}^{10}y_{i}^j log(h_i^j)+(1-y_{i}^j)log(1-h_i^j)\\
h_i^j=\frac{o_{i}^j}{\sum_{n=1}^{10}o_{n}^j}
\end{gathered}
\end{equation}

where $j$ represents the $j^{th}$ image in the batch. From here, the derivation of the gradient can commence.

\begin{equation}\label{eq_s5}
\begin{gathered}
\frac{dC}{d\varepsilon_{(r)}}=\sum_{j=1}^m \bigg(\int\frac{\partial C}{\partial E^j_{(r')}}\frac{\partial E^j_{(r')}}{\partial\varepsilon_{(r)}}\,dr' + \int\frac{\partial C}{\partial \overline{E^j_{(r')}} }\frac{\partial \overline{E^j_{(r')}}}{\partial\varepsilon_{(r)}}\,dr'\bigg)\\
\end{gathered}
\end{equation} 

where $\overline{E^j_{(r')}}$ is the conjugate of the electric field. Here directly computing the above gradient involves calculation of $\partial E^j_{(r')}/\partial\varepsilon_{(r)}$, which is computationally expensive. In order to circumvent this problem, we define a Lagrangian as 

\begin{equation}\label{eq_s6}
Lg=C+\sum_{j=1}^m \int\lambda^j_{(r')}(L_{(r')}E^j_{(r')}+i\omega J^j)\,dr'+\sum_{j=1}^m \int\overline{\lambda^j_{(r')}(L_{(r')}E^j_{(r')}+i\omega J^j)}\,dr'
\end{equation}

Here $\lambda^j_{(r')}$ is the Lagrange multiplier which is equivalent to \textit{Adjoint Field} in similar works. This Lagrangian has the same gradient as the cost function because the second and third terms are zero. We now compute its gradient with respect to the real part of the permittivity at each spatial point:

\begin{equation}\label{eq_s7}
\begin{gathered}
\frac{dLg}{d\varepsilon_{(r)}}=
\int\sum_{j=1}^m \frac{\partial C}{\partial E_{(r')}^j}\frac{\partial E_{(r')}^j}{\partial\varepsilon_{(r)}}\,dr'+\int\sum_{j=1}^m \frac{\partial C}{\partial \overline{E_{(r')}^j}}\frac{\partial \overline{E_{(r')}^j}}{\partial\varepsilon_{(r)}}\,dr'+\\
\int\sum_{j=1}^m \lambda^j_{(r')}(\frac{dL_{(r')}}{d\varepsilon_{(r)}}E^j_{(r')}+L_{(r')}\frac{\partial E^j_{(r')}}{\partial\varepsilon_{(r)}})\,dr'+
\int\sum_{j=1}^m \overline{\lambda^j_{(r')}}(\frac{d\overline{L_{(r')}}}{d\varepsilon_{(r)}}\overline{E^j_{(r')}}+\overline{L_{(r')}}\frac{\partial \overline{E^j_{(r')}}}{\partial\varepsilon_{(r)}})\,dr'
\end{gathered}
\end{equation}

In Eq.S7, since the dielectric constant of nonlinear materials depends on the field, ${dL_{(r')}}/{d\varepsilon_{(r)}}$ can be calculated as 

\begin{equation}\label{eq_s8}
\frac{dL_{(r')}}{d\varepsilon_{(r)}}=\frac{\partial L_{(r')}}{\partial\varepsilon_{(r)}}+\frac{\partial L_{(r')}}{\partial E^j_{(r')}}\frac{\partial E^j_{(r')}}{\partial\varepsilon_{(r)}}+\frac{\partial L_{(r')}}{\partial \overline{E^j_{(r')}}}\frac{\partial\overline{E^j_{(r')}}}{\partial\varepsilon_{(r)}}
\end{equation}

Similarly, ${d\overline{L_{(r')}}}/{d\varepsilon_{(r)}}$ can be calculated too. In order to apply the adjoint state method, we first group all the terms multiplied by ${\partial E_{(r')}^j}/{\partial\varepsilon_{(r)}}$ together (likewise for ${\partial \overline{E_{(r')}^j}}/{\partial\varepsilon_{(r)}}$). This provides us with Eq.S9.

\begin{equation}\label{eq_s9}
\begin{gathered}
\frac{dLg}{d\varepsilon_{(r)}}=2Real\{\int\sum_{j=1}^m \bigg(\frac{\partial C}{\partial E_{(r')}^j}+
\lambda^j_{(r')}(L_{(r')}+E^j_{(r')}\,\frac{\partial L_{(r')}}{\partial E^j_{(r')}})+\overline{\lambda^j_{(r')}}(\overline{E^j_{(r')}}\,\frac{\partial \overline{L_{(r')}}}{\partial E^j_{(r')}})\bigg)\frac{\partial E_{(r')}^j}{\partial\varepsilon_{(r)}}\,dr'\}+\\
2Real\{\int\sum_{j=1}^m \lambda^j_{(r')}\frac{dL_{(r')}}{d\varepsilon_{(r)}}E^j_{(r')}\,dr'\}
\end{gathered}
\end{equation}

we set the Lagrange multiplier $\lambda^j$ in a manner that forces the summation of all the terms multiplied by ${\partial E_{(r')}^j}/{\partial\varepsilon_{(r)}}$ to be zero. $\lambda^j$ can be calculated with the \textit{Adjoint Equation} that is produced in the equation above by setting the first term in it to zero.

\begin{equation}\label{eq_s10}
\frac{\partial C}{\partial E_{(r')}^j}+\lambda^j_{(r')}(L_{(r')}+E^j_{(r')}\frac{\partial L_{(r')}}{\partial E^j_{(r')}})+\overline{\lambda^j_{(r')}}(\overline{E^j_{(r')}}\frac{\partial \overline{L_{(r')}}}{\partial E^j_{(r')}})=0
\end{equation}

When the adjoint equation is satisfied,  the gradient can be further simplified as :

\begin{equation}\label{eq_s11}
\frac{dLg}{d\varepsilon_{(r)}}=2Real\{\int\sum_{j=1}^m \lambda^j_{(r')}\frac{dL_{(r')}}{d\varepsilon_{(r)}}E^j_{(r')}\,dr')\}
\end{equation}

Using the definition of the operator $L_{(r')}$ in Eq.S1, we get $\frac{dL_{(r')}}{d\varepsilon_{(r)}} =-\omega^2\frac{d\varepsilon_{(r')}}{d\varepsilon_{(r)}}=-\omega^2\delta_{(r,r')}$, which brings us to the following equation.

\begin{equation}\label{eq_s12}
\frac{dLg}{d\varepsilon_{(r)}}=-2\omega^2\,Real\{\sum_{j=1}^m\lambda^j_{(r)} E^j_{(r)}\}
\end{equation}

The next step in deriving the gradient is calculating the derivative of cost with respect to the electric field.

\begin{equation}\label{eq_s13}
\frac{\partial C}{\partial E_{(r)}^j} = -\frac{1}{m}\sum_{i=1}^{10} y_i^j\frac{1}{h_i^j}\frac{\partial h_i^j}{\partial E_{(r)}}-(1-y_i^j)\frac{1}{1-h_i^j}\frac{\partial h_i^j}{\partial E_{(r)}}
\end{equation}

The term ${\partial h_i^j}/{\partial E_{(r)}}$ can be calculated based on the definition of $h_i^j$ in Eq.S4.

\begin{equation}\label{eq_s14}
\frac{\partial h_i^j}{\partial E_{(r)}}=\frac{\frac{\partial o_i^j}{\partial E_{(r)}}(\sum_{k=1}^{10} o_k^j)-o_i^j(\sum_{k=1}^{10}\frac{\partial o_k^j}{\partial E^j_{(r)}})}{\sum_{k=1}^n o_k^j}
\end{equation}

In addition, we can easily obtain

\begin{equation}\label{eq_s15}
\begin{gathered}
\frac{\partial o_k^j}{E^j_{(r)}}=\frac{\varepsilon_{air}}{2}\frac{\partial}{E^j_{(r)}}\int E_{(r')}^j\overline{E_{(r')}^j}R_{k(r')}dr'=\frac{\varepsilon_{air}}{2}\int\delta_{(r,r')}\overline{E_{(r')}^j} R_{k(r')}dr'=\frac{\varepsilon_{air}}{2}\overline{E_{(r)}^j}R_{k(r)}\\
\Rightarrow\frac{\partial C}{\partial E_{(r)}^j} = -\frac{\varepsilon_{air}}{2m}\sum_{i=1}^{10}\frac{y_i^j-h_i^j}{o_i^j(1-h_i^j)}(R_{i(r)}-h_i^j\sum_{k=1}^{10}R_{k(r)})\overline{E_{(r)}^j}
\end{gathered}
\end{equation}

\subsection{Nonlinear Materials}

The nonlinearity that we have implemented in this work was designed to loosely follow the behavior of a ReLU activation function\cite{nair2010rectified}. Of course, applying this activation function directly to the field would not be meaningful, so it had to be modified to act upon the intensity of the field. The idea here was to design the nonlinearity so that fields with intensity below a certain threshold are blocked. This objective can be achieved by adding an imaginary part of the form $-{\alpha}/{|E|^2}$  to the permittivity.

\begin{figure}[ht]
\centering
\includegraphics[scale=1]{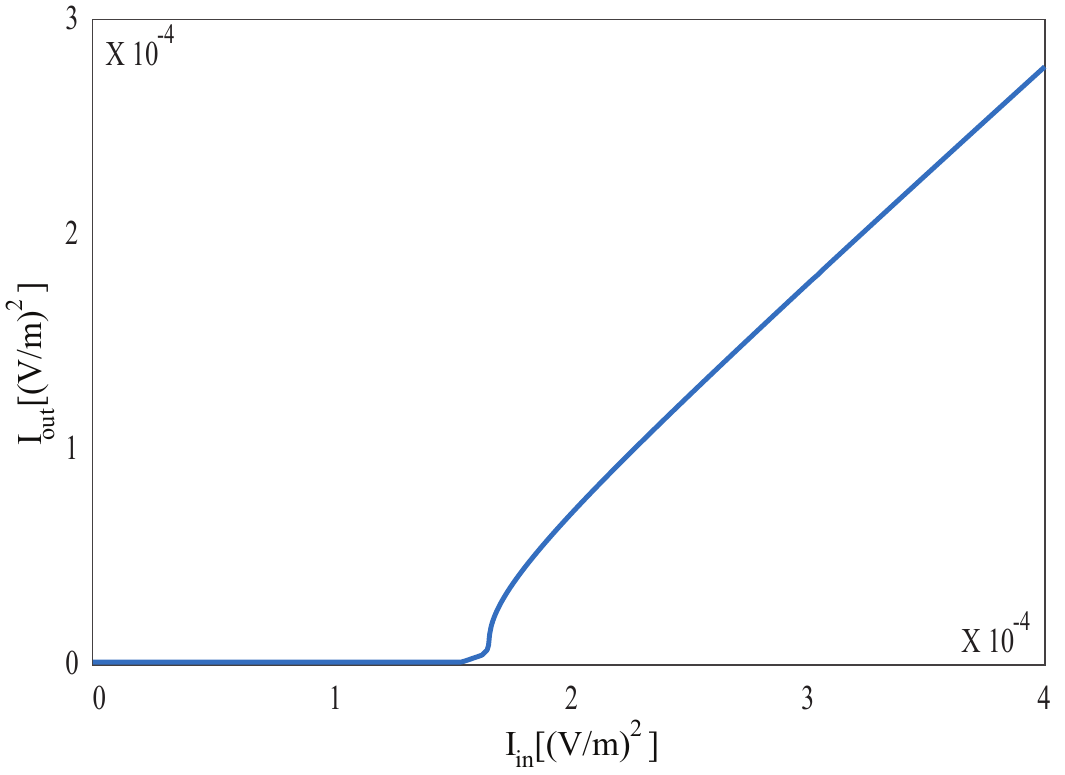}
\caption{The intensity response of the nonlinear layer in 1D as a function of the input intensity for a length of $\lambda/2$ and an $\alpha$ equal to $5\times 10^{-5}\,(\frac{V}{m})^2$.}.
\label{fig:nl_graph}
\end{figure}

Although this type of nonlinearity was chosen rather arbitrarily, it alters the intensity in a similar way as optical saturable absorbers\cite{shen2017deep}. With the details of the proposed nonlinearity explained, its contribution to the gradient can now be assessed. To that end, we first define the permittivity in nonlinear sections as $\varepsilon_{(r)}=\varepsilon_{real(r)}+i\varepsilon_{img(r)}$ where $\varepsilon_{real(r)}$ and $\varepsilon_{img(r)}$ are the real and imaginary parts of the permittivity respectively. As was shown earlier, the gradient has an extra term in nonlinear sections of the structure. The extra terms can now be calculated for this particular nonlinearity. The first extra term is calculated as follows

\begin{equation}\label{eq_s16}
\begin{gathered}
\frac{\partial L_{(r)}}{\partial E^j_{(r)}}=\frac{\partial L_{(r)}}{\partial\varepsilon_{(r)}}\frac{\partial\varepsilon_{(r)}}{\partial E^j_{(r)}}=(\omega^2)(-i\alpha\frac{\overline{E^j_{(r)}}}{|E^j_{(r)}|^4})=i\omega^2\varepsilon_{img(r)}\frac{1}{E^j_{(r)}}\\
L_{(r)}+E^j_{(r)}\frac{\partial L_{(r)}}{\partial E^j_{(r)}}=(\nabla\times(\frac{1}{\mu})\times\nabla\times)-\omega^2\varepsilon_{(r)}+i\omega^2\varepsilon_{img(r)}=(\nabla\times(\frac{1}{\mu})\times\nabla\times)-\omega^2\varepsilon_{real(r)}
\end{gathered}
\end{equation}

While the final term takes the following form.

\begin{equation}\label{eq_s17}
\begin{gathered}
\overline{E^j_{(r)}}\frac{\partial \overline{L_{(r)}}}{\partial E^j_{(r)}}=\overline{E^j_{(r)}}\frac{\partial \overline{L_{(r)}}}{\partial\,\overline{\varepsilon_{(r)}}}\frac{\partial\,\overline{\varepsilon_{(r)}}}{\partial E^j_{(r)}}=\omega^2i\alpha\frac{\overline{E^j_{(r)}}\,\overline{E^j_{(r)}}}{|E^j_{(r)}|^4}
\end{gathered}
\end{equation}

The final tool we need before putting all the concepts together is a way to solve the nonlinear wave equation. We use Finite-Differences Frequency-Domain(FDFD) method. In computational domain the wave equation can be written as a matrix multiplication shown in Eq.S18. 

\begin{equation} \label{eq_s18}
AE=-i\omega{J}
\end{equation}

In this equation $A=(C_{h}D_{\mu}^{-1}C_{e}-\omega^2{D_{\varepsilon}})$, where $C_{h}$ and $C_{e}$ are curl matrices, $D_{\mu}$ and $D_{\varepsilon}$ represent permeability and permittivity of different points in the domain respectively; while $J$ represents the current source density matrix in the computational domain. Here we use iterative method to solve the nonlinear response. Finally, for the implementation of this method, we utilized the MATLAB-based package MaxwellFDFD\cite{maxwellfdfd-webpage}. At this stage, with all the necessary pieces available, we can finally formulate the problem in the forward and backward directions. Solving for the electric field in the forward direction is of this form

\begin{equation}\label{eq_s19}
\begin{gathered}
E^j=(A^j)^{-1}b^j\\
\frac{\partial C}{\partial \varepsilon} = -2\omega^2 real\{\sum_{j=1}^m \lambda^j\odot E^j\}\\
\end{gathered}
\end{equation}

While the adjoint equation (the problem of the adjoint field propagating backwards) takes the following form.

\begin{equation}\label{eq_s20}
\begin{gathered}
\frac{-\varepsilon}{m} \Big(\big(\frac{y_{i}^j-h_i^j}{o_{i}^j(1-o_{i}^j)}\big) \big(\gamma_{i} - h_i^j \sum_{n=1}^{10}\gamma_{n}\big)\odot \overline{E^j}\Big)+\lambda^j(A^j-\omega^2i\alpha\frac{1}{E^j\odot \overline{E^j}})+\overline{\lambda^j}(\omega^2i\alpha\frac{1}{E^j\odot E^j})=0
\end{gathered}
\end{equation}

In the above equations $\odot$ represents the Hadamard product(the element wise multiplication between two matrices). Eq.S20 enables us to calculate the gradient with respect to each of the elements in the computational domain all at once. Once this gradient is computed, it is possible to choose which elements to update. Of course, since updating the permittivity of the nonlinear sections would lead to a medium that is far from any real physical interpretation, we decided to restrict the updating process to the linear sections of the medium.

\subsection{Level-Set Evolution}
As mentioned in the main paper, updating the structure at each step is done by evolving the boundary between the two constituent materials with the following equation.

\begin{equation} \label{eq_s26}
\partial_t\phi+v(x,y)|\nabla\phi|=0
\end{equation}

Where $\phi$ is the level-set function and $v(x,y)$ is the velocity with which each point on the zero crossing curve of the level-set function moves normal to the the curve(which is set equal to the gradient with respect to permittivity of different points in our work). However, simply implementing this equation in the digital domain does not maintain a stable curve evolution process. To overcome this issue,  we implemented the method introduced in ref \cite{li2010distance} where an extra term is added to the evolution equation to ensure the level-set function remains a signed distance function through its evolution.

\begin{equation} \label{eq_s27}
\begin{gathered}
\partial_t\phi=\mu \nabla.\big(p(|\nabla \phi|)\nabla\phi\big) - v(x,y)|\nabla\phi|\\
p(s)=\left\{
                \begin{array}{ll}
                  \frac{1}{(2\pi)s}sin(2\pi s)\quad s\leq 1\\
                  	1-\frac{1}{s}\qquad\quad\quad s\geq 1
                \end{array}
              \right.
\end{gathered}
\end{equation}

Where $\mu$ is a constant(set to $0.2$ in this work) and $p(s)$ is a double-well potential for distance regularization. With this approach, we do not need to concern ourselves with matters of stability for the level-set function.

\subsection{Training with FDFD Simulation}

As discussed in the paper, both a 2D and a 3D model with the proposed method were trained. This training process required solving the wave equation for different inputs and in this section the details the FDFD simulations and training process are explained.

\begin{figure}
\centering
\includegraphics[scale=1]{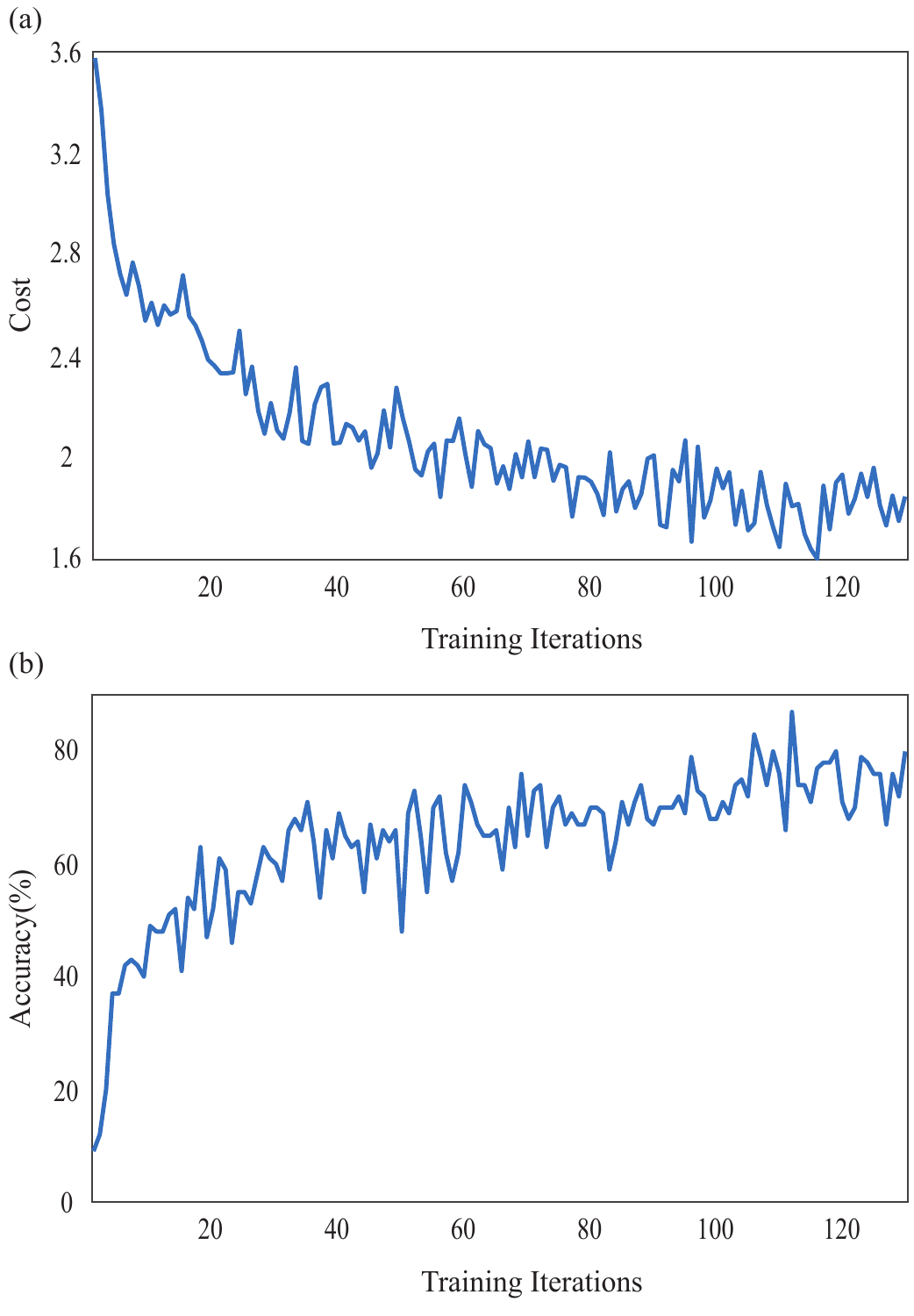}
\caption{(a) The training curve for 2D NNM. (b) Evolution of the accuracy with each training iteration on batches of test set equal in size to the training batch for the same medium.}
\label{fig:loss_curve}
\end{figure}

The 2D version of the medium was trained in a domain with PML boundary condition on all sides, while the 3D simulations were done with the PEC boundary condition on planes parallel to the direction of wave propagation and PML in the direction perpendicular to it. The different boundary conditions for the 3D version were implemented to reduce the computational load of the simulation. The wavelength for both simulations was set to $1\mu m$, while the spatial resolution of the simulation was set to $\lambda/20$ and $\lambda/5$ for 2D and 3D versions respectively.

In the main manuscript we went on to explain how the iterative process of solving the nonlinear wave equation worked. We explained that we start with a random initial wave distribution that effectively gave us the initial permittivities, and then use these initial values to solve for the new field distribution. This process was then repeated until the electric field converged to a fixed distribution. However, in practice, we start with a field solution corresponding to one of our input images. This helps the field to converge more quickly .

As was mentioned in the paper, we used SGD for the training process. More precisely we used a variation of SGD called mini-batch gradient descent. While SGD uses all the input data to calculate the gradient for each training iteration, mini-batch gradient descent only uses a small batch of the input data. This significantly increases the convergence speed of the optimization process. Of course, the use of batches was already hinted at in the main text, so we continue by explaining some of the implementation details regarding batches and learning rate. We have used batches of 100 for the 2D training and batches containing 50 samples for the 3d version. The larger batch size for the 2D version was chosen so, because the selected method of training showed that it was more sensitive to batch size and therefore it was better to train with a larger batch. Finally, the learning rate for the 2D version was set to 500 which was reduced after two epochs of training, while its 3D counterpart was set to 30.

Fig.S3 shows the cost values for the training set and the accuracy values for the test set on the 2D binary medium. 

In the main text, we showed the confusion matrix for the case of the 3D NNM. On one axis, this matrix shows the ground truth labels for each input image, and on the other, the labels produced by the NNM for that image. Therefore, the diagonal elements show the number of correct classifications by the NNM, while the rest of the elements on each row show the wrong classifications. In Fig.S5, the confusion matrix for the 2D NNM has been depicted where 10 samples of each digit have been presented for the medium.

\begin{figure}[ht]
\centering
\includegraphics[scale = 1.5]{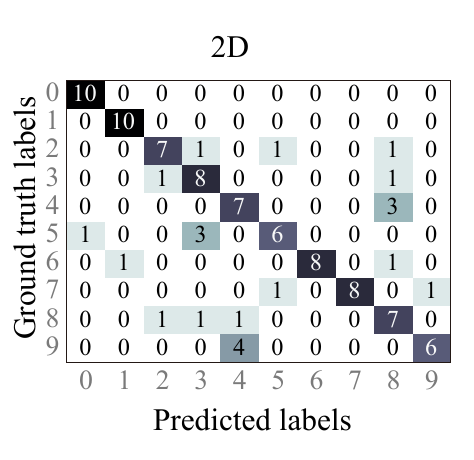}
\caption{Confusion matrix for the 2D PNCM. The diagonal elements show the number of correct classifications among the 10 input digits presented as the input, while the other columns show what the wrong classes that each digit was categorized as are.}
\label{fig:confu2d}
\end{figure}

\section*{Funding}
The work was financially supported by DARPA Young Faculty Award program.

\section*{Acknowledgments}
The authors thank W. Shin for his help on improving the computational speed of the implementation of this method.

\section*{Disclosures}
The authors declare that there are no conflicts of interest related to this article.

\section*{Conclusion}
Here we show that the wave dynamics in the Maxwell’s equations is capable of performing highly sophisticated computing. There is intricate connection between differential equations that govern many physical phenomena and neural computing (see more discussion in supplementary), which could be further explored. From the perspective of optics, the functions of most nanophotonic devices can be described as mode mapping \cite{miller2012all}. In traditional nanophotonic devices, mode mapping mostly occurs between eigenmodes. For example, a polarization beam splitter \cite{shen2015integrated} maps each polarization eigenmode to a spatial eigenmode. Here, we introduce a class of nanophotonic media that can perform complex and nonlinear mode mapping equivalent to artificial neural computing. The neural computing media shown here has an appearance of disorder media. It would be also interesting to see how disorder media, which support rich physics such as Anderson localization, could provide a new platform for neural computing. In comparison, today's  optical  neural  computing mostly follows layered  structures. While  highly  efficient  for  digital  computing,  layered  structures could be counter-productive in optical analog computing. For  practical  applications that routinely  use  millions  of connections, it  can be challenging to implement deep and dense layers of optical components in a compact form. Nevertheless, the concept of NNM shows that any nanostructures could be optimized to perform neural computing without the rigid constraint of layer structures. Combined with ultra-high computing density, NNM could be used in a wide range of information devices as the analog preprocessing unit.


\bibliography{bibl}

\end{document}